\documentclass[aps,pre,reprint,showpacs,superscriptaddress,groupaddress,showkeys,floatfix,byrevtex, bibnotes]{revtex4-1}
\usepackage[colorinlistoftodos]{todonotes}

\usepackage{graphicx}
\usepackage{slashed}
\usepackage{amsfonts}
\usepackage{amsmath}
\usepackage{color}
\usepackage{soul}
\usepackage{tensor}
\usepackage{srctex}
\usepackage{dsfont}
\usepackage{amsmath}
\usepackage{natbib}

\begin{document}

\title{Active motion on curved surfaces
}

\author{Pavel \surname{Castro-Villarreal}}
\email[]{pcastrov@unach.mx}
\thanks{author to whom correspondence should be addressed.}
\affiliation{Facultad de Ciencias en F\'isica y Matem\'aticas,
Universidad Aut\'onoma de Chiapas, Carretera Emiliano Zapata, Km. 8, Rancho San Francisco, 29050 Tuxtla Guti\'errez, Chiapas, M\'exico}

\author{Francisco J. \surname{Sevilla}}
\email[]{fjsevilla@fisica.unam.mx}
\thanks{author to whom correspondence should be addressed.}
\affiliation{Instituto de F\'isica, Universidad Nacional Aut\'onoma de M\'exico,
Apdo.\ Postal 20-364, 01000, M\'exico D.F., Mexico}

\begin{abstract}
A theoretical analysis of active motion on curved surfaces is presented in terms of a generalization of the Telegrapher's equation. Such generalized equation is explicitly derived as the polar approximation of the hierarchy of equations obtained from the corresponding Fokker-Planck equation of active particles diffusing on curved surfaces. The general solution to the generalized telegrapher's equation is given for a pulse with vanishing current as initial data. Expressions for the probability density and the mean squared geodesic-displacement are given in the limit of weak curvature. As an explicit example of the formulated theory, the case of active motion on the sphere is presented, where oscillations observed in the mean squared geodesic-displacement are explained.

\end{abstract}

\pacs{}
\keywords{Active Particles, Curved Manifolds, Telegrapher's equation}

\maketitle

\section{Introduction} 

{\it Active matter} is the term coined to those systems in soft condensed matter, that find themselves in out of equilibrium conditions. These systems are composed by \emph{self-propelled particles} that are capable of converting the locally absorbed energy from their environment into motion. On the one hand, there have been a great of interest in the collective properties that emerge in these out-of-equilibrium systems. For instance, alignment interactions between self-propelled particles (also known as \emph{active particles}), give rise to wonderful patterns describing systems such as a flock of birds, ants and bacterial colonies, schools of fishes \cite{VicsekPhysRep2012}, even non-living matter like thermal active colloids \cite{GolestanianPRL2012}, among others. Interesting, out-of-equilibrium, collective phenomena, emerge even from simple  alignment interactions as those originally considered by Vicsek {\it et al.} \cite{VicsekPRL1995}, and described theoretically by the used of continuum field theories by Toner and Tu in the 90's \cite{TonerPRL1995}, ideas that have been developed extensively during the last few decades \cite{RamaswamyAnnRevCondMattPhys2010,MarchettiRMP2013}. Recently, the inclusion of spatial and alignment interactions among active particles were considered \cite{FarrelPRL2012}, while the complete phase diagram of the Vicsek {\it et al.} model has been recently interpreted in terms of a gas-liquid transition \cite{SolonPRE2015}. Furthermore, it has been pointed out that self-propulsion is not a necessary intrinsic property in order to explain collective motion \cite{DossettiPRL2015}.

One of the topics, that is being recently developed, is the active dynamics in heterogeneous situations, for example, when the active dynamics occurs or is limited by a curved surface.
It turns out that processes, for which active matter systems are confined to a curved surface, abound in biology.
For instance, the embryonic developmental processes can be thought as a collective cellular movement controlled by a curved surface, the embryonic sac; the cell movement on the development of a corneal growing; the transport of blood cell through the vascular system; the flocking motion od species that migrates into different regions of the earth; etc. As a result, the importance of the effects, due to geometric and/or topological features of the embedding space, on the diffusion of active particles has been emphasized. 

In a broad way, it is generally possible to distinguish the local effects on the transport properties of active particles due to the curvature of the embedding manifold from 
the global ones due to its topology. 
This means, in particular, that the curvature effects will be revealed locally in a neighborhood of a given point, specially, in the short-time regime of the particle 
dynamics. Topological effects on the other hand, will be manifested in the opposite limit, i.e. in the long-time regime. The curvature effects have been observed in the 
dynamics of self-propelled particles inside curved (convex and non-convex) walls, where the probability density function depends strongly on the curvature of the confining 
surface \cite{FilySoftMatt2014,FilyPRE2015,FilyEPJE2017}. Subsequently, these ideas were generalized to the case for which the active dynamics of interacting particles occurs 
on a curved convex surface \cite{Fily2016}. 
In addition, the geometry of obstacles or micro-components, immersed in active fluids, induces an active behavior in its own dynamics \cite{MalloryPRE2014}. In another study, 
guided by a semi-empirical proposal, the swimming pressure was determined on curved and sharp rigid walls \cite{SmallenburgPRE2015} and more recently for soft walls 
\cite{NikolaPRL2016}. Further, it has been shown that vortex motion appeared in an active polar fluid confined to a curved surface like  ellipsoid is related to the umbilical 
points of the same surface \cite{EhrigPRE2017} and that a spontaneous active flow appears when nematic active fluid is distorted by a curved substrate \cite{Green2016}. 

On the other hand, the effects of topology have been pinpointed in different recent studies. In Ref. \cite{KeberScience2014} the emergence of spatiotemporal patterns is analyzed when the active microtubules and molecular motors are encapsulated in a lipid vesicle. It turns out that the topology of the sphere is determinant for the formation of oscillating patterns characterized by the dynamics of topological defects in the sphere. Also, the emergence of movement patterns of the polar vortex and circulating band were observed as a consequence of the spherical topology \cite{SknepnekPRE2015, Henkes2017}. Recently, on the basis of the extended theory of Toner and Tu hydrodynamics to curved surfaces, it has been found that the curvature induces a gap in the acoustic spectrum which is associated with topologically protected modes \cite{ShankarPRX2017}.

On the general theoretical setting, the overdamped kinematics of an active Brownian particle is described by a pair of Langevin-like equations, one for the particle’s position and the other for the swimming direction \cite{VicsekPhysRep2012}. A coarse-grained description in terms of the position alone leads to Smoluchowski-like equation which, in the open Euclidean space and in the long-time description of the active particle dynamics, it is approximately described by the so-called telegrapher equation (TE) \cite{SevillaPRE2014,SevillaPRE2015,SevillaPRE2016}. Such an equation accounts for persistent Brownian motion if coherent initial distributions are avoided in order to maintain the positiveness of the probability density \cite{PorraPRE1997}, i.e., it is valid  in the diffusive regime. Notwithstanding this, it has been proved, that the TE provides the whole exact time dependence of the mean squared displacement of a self-propelled particle \cite{SevillaPRE2014,SevillaPRE2015}, and a good approximation for the time dependence of higher moments valid only in the long-time regime.

The TE has been widely studied since it describes transport phenomena in many different contexts. In the one-dimensional straight-line, the TE was introduced in the middle of 
the last century to take into account the effects of persistence in the trajectories of random walkers \cite{GoldsteinQJMAM1951}, a model the considers the effects of the 
finite speed propagation of the particles \cite{DunkelPhysRep2009}. 
The straightforward generalization of the telegrapher equation to dimensions larger than one, has interpretation problems of the dynamical quantity of interest, to say, of the 
probability density of the diffusion process of persistent walkers, since its solutions breaks down the positiveness of the probability density in the short-time regime as 
pointed out in Ref. \cite{PorraPRE1997, GodoyPRE1997}. The origin of this feature can be traced back to the wake effects of the wave-like behavior of the solutions in that 
specific regime. This contrasts with the situation of an ensemble of non-interacting, persistent random walkers that move in continuous, two-dimensional space 
\cite{SevillaPRE2014}, for which it is clear that for times shorter than the persistence time, no wake effects are apparent leading to a sharp front of particles that move 
almost ballistically. 
No doubt that these issues also appear in the telegrapher equation defined on Riemannian manifolds. However, in both open Euclidean and Riemannian spaces there is an appropriate parameter region where probability density is positive \cite{PorraPRE1997, SevillaPRE2014}.

In this work, a theory of stochastic Langevin equations is proposed to describe the kinematic state of a $2D$ active particle confined to move on a curved surface. After 
deriving the Fokker-Planck equation, we develop a hydrodynamic description for the ideal active gas on the surface. This description is given through a hierarchy of field 
equations for the hydrodynamic tensors i.e. particle density $\rho$, polarization field $\mathbb{P}^{a}$, nematic tensor field $\mathbb{Q}^{ab}$, etc. \cite{FarrelPRL2012}.  
One of the main hypothesis we use is the polar approximation which consist of consider $\mathbb{Q}_{ab}$ a fast variable and homogeneously distributed over the points of the 
surface. In particular, we are able to prove that within this approximation $\rho$ satisfies the TE as it occurs in the flat case \cite{SevillaPRE2014}. The general covariance 
of the resulting TE is exploited in order to study the curvature effects appearing in the particle density, and particularly in the mean squared geodesic-displacement (MSGD). 
The calculation are performed using the local frame provide by the Riemann Normal Coordinates (RNCs) \cite{EisenhartBook}. 
Finally,  we explore the active system confined to the sphere where we are able to confirm the general results. 
 
This paper is organized as follows. In section \ref{SectII}, we present the Langevin equations of motion for an active particle confined to a curved surface. Also, the Fokker-Planck equation, associated to the stochastic equations, is derived in order to build a standard hydrodynamics description for the ideal active gas. In section \ref{SectIII} we discuss the general aspects regarding the evolution of the particle density on the surface, when the polar approximation is applied. In section \ref{SectIV},  we provide analytical expressions in the limit of weak curvature. In addition, the mean-square displacement is studied under the same circumstances. In the section \ref{SectV}, we provide a recurrence equation that provide the complete behavior of the active particles on the sphere. Additionally, in the polar approximation we are able to test our general results of the two previous sections. Finally, in the section \ref{SectVI}, we give our concluding remarks and perspectives of this work.   

\section{\label{SectII} Active motion on curved surfaces}

The kinematic state of an active particle that swims with constant speed $v_{0}$ on a two-dimensional curved surface $S$ is determined by 
its position $\boldsymbol{x}(t)$ and the direction of motion $\hat{\boldsymbol{v}}_{\text{swim}}(t)$. The particle position on the surface is described by the set of local 
coordinates $\{x^{a}\}$ with $a=1,2$, while, the self-propulsion director $\hat{\boldsymbol{v}}_{\text{swim}}(t)$, that changes the position of the particle along the surface, 
is contained only on the tangential plane to the surface at the point where the particle is located, i.e. the dynamics of this quantity is two dimensional and is described by 
the evolution of the coordinates $v_{\text{swim}}^{a}$, which must
satisfy the condition $v_{\text{swim}}^{a}\left(t\right)v_{\text{swim}}^{b}\left(t\right)g_{ab}\left[x(t)\right]=1$, where $g_{ab}$ is the me\-tric tensor that characterizes 
the Riemannian geometry of the surface. The stochastic equations that give the dynamics of the active particle on a curved surface are (see appendix \ref{appendix1} for a 
derivation)
\begin{subequations}
\label{eq1}
\begin{align}
\frac{d}{dt}x^{a}(t)=&\, v_{0}v_{\text{swim}}^{a}(t),
\label{eq1a}\\
\frac{d}{dt}v_{\text{swim}}^{a}(t)=&-v_{0}\, v_{\text{swim}}^{c}(t)v_{\text{swim}}^{d}(t)\Gamma\indices{^{a}_{cd}}\, +\nonumber\\
&\qquad\sqrt{2\gamma\, g[x(t)]}\, v_{\text{swim}}^{c}(t)\, \epsilon_{cd}\, g^{da}\, \zeta[x(t)],
\label{eq1b}
\end{align}
\end{subequations}
where Einstein notation, summation over repeated indexes, is assumed; $\gamma$ is the strength of the active fluctuations that affects the direction of motion. Notice that with $\gamma$ that has units of time$^{-1}$ and $v_{0}$, the characteristic length scale, $L=v_{0}/\gamma$, measures the average distance that an active particle travels along a given direction and is called the \emph{persistent length}. $\zeta[x(t)]$ is a scalar noise that depends explicitly on the particle state through the projection of $\boldsymbol{\zeta}(t)$, a three-dimensional vector whose entries are Gaussian white noise with zero mean, along 
${\bf N}[x^{a}(t)]$ the local, unitary, normal vector to the tangential plane located at the particle position $x^{a}(t)$ on the surface, to say $\boldsymbol{\zeta}(t)\cdot {\bf N}[x^{a}(t)]$. $g^{ab}$ and $\epsilon_{ab}$ denote the inverse of the metric tensor and the two dimensional Levi-Civita tensor, respectively, while $\Gamma\indices{^{a}_{cd}}$ denotes the Christoffel symbols. 
All these tensors encompasses the geometrical data of the intrinsically curved surface. Particularly, for a flat surface, where one has $\Gamma\indices{^{a}_{cd}}=0$ and 
$g_{ab}=\delta_{ab}$, the above stochastic equations (\ref{eq1}) reduce to the equations studied in Ref. \cite{SevillaPRE2014}. An slightly different approach, that considers 
the interaction between active particles, confined on a curved surface is presented in \cite{Fily2016}.  

From the above stochastic equations (\ref{eq1}), one can get the Fokker-Planck equation for the one-particle distribution function
\begin{eqnarray}
P(x, v, t)=\left<\frac{1}{\sqrt{g\left(x\right)}}\delta\left[x^{a}-x^{a}\left(t\right)\right]\delta\left[v^{a}-v_{\text{swim}}^{a}\left(t\right)\right]\right>,\nonumber
\end{eqnarray}
where $\left<\cdots\right>$ denotes the average over the realizations of the active fluctuations $\zeta[x(t)]$ and $g(x)=\det[g_{ab}(x)]$. Strictly speaking, $P(x, v,t)$ depends also on the initial values of the position and velocity. 

By following standard methods \cite{Zinn-JustinBook}, it is straightforward to show that the Fokker-Planck that corresponds to the Eqs. \eqref{eq1} is
\begin{multline}\label{Fokker-Planck}
\frac{\partial P}{\partial t}=\gamma\frac{\partial}{\partial v^{a}}\left[\left(g^{ab}v^2-v^{a}v^{b}\right)\frac{\partial P}{\partial v^{b}}\right]-\nabla_{a}\left(v_{0}v^{a}P\right)\\
+\frac{\partial}{\partial v^{a}}\left(v_{0}v^{c}v^{d}\Gamma\indices{^{a}_{cd}}P\right),
\end{multline}
where $v^{2}=v^{a}v^{b}g_{ab}$ and $\nabla_{a}$ denotes the covariant derivative compatible with the metric $g_{ab}$. Last equation takes into account, explicitly, the effects 
of the surface curvature on the dynamics of an active particle constraint to move on that surface. The first term accounts for the internal fluctuations on the direction of 
motion, which occurs on the  tangent plane located at particle location on the surface. The second term gives the flux due to the self-propulsion on the surface. The third 
term, accounts for the constrained motion on the surface as is evidenced by the appearance of the Christoffel symbols.

The coupling of the coordinates $x^{a}$, through the derivatives, with the variables $v^{a}$ in the second term of the right hand side in equation \eqref{Fokker-Planck}, hinders the obtaintion of an exact solution, however, a thorough analysis can be carried out along different methods \cite{CatesEPL2013, SchnitzerPRE1993, SevillaPRE2014, SevillaPRE2015}. We follow the standard coarse-graining procedure to give a stochastic-like hydrodynamic description of Eq. \eqref{Fokker-Planck} \cite{DuderstadtTransportTheory,CatesEPL2013,Fily2016}, namely, we consider the hierarchy of equations that relates the probability density $\rho(x,t\vert x^{\prime})=\int d^{2}v ~P(x, v, t)$; the polarization field $\mathbb{P}^{a}\left(x,t\right)=\int d^{2}v ~v^{a} P(x,v,t)$; the nematic order parameter $\mathbb{Q}^{ab}\left(x,t\right)=\int d^2v (v^{a}v^{b}-\frac{1}{2}g^{ab})P(x,v,t)$ and higher order hydrodynamic tensors. 
Notice that, these tensor fields $\rho(x,t\vert x^{\prime})$, $\mathbb{P}^{a}(x,t)$, $\mathbb{Q}^{ab}(x,t)$, $\ldots$, can be put in univocally correspondence with a standard 
multipole expansion of the one-particle  distribution function $P(x, v, t)$. Some symmetries are evidenced in these quantities for instance, it can be proved that nematic 
order parameter is traceless with respect to the metric tensor, i.e. $g_{ab}\mathbb{Q}^{ab}=0$. In the following, we briefly depict how the hierarchy of equations for these 
hydrodynamic quantities emerge from \eqref{Fokker-Planck} \citet{Fily2016}. 

After a first integration over the velocity domain on the both sides of the Fokker-Planck equation, the only term that survives from Eq. (\ref{Fokker-Planck}) is the middle one, and therefore we get the continuity-like equation
\begin{eqnarray}
\frac{\partial \rho}{\partial t}=-\nabla_{a}\left(v_{0}\mathbb{P}^{a}\right),
\label{cont}
\end{eqnarray}
where $v_{0}\mathbb{P}^{a}$ can be interpreted as the probability current on the surface. Now, we need the evolution equation for the polarization field $\mathbb{P}^{a}(x,t)$. 
We should determine this equation  carefully  since covariant derivative acts on the rank-$2$ tensors in a particular way (different than how it does on vectors) 
\cite{NakaharaBook}; this is the reason how the Christoffel symbols disappear explicitly in the equation for $\mathbb{P}^{a}$
\begin{eqnarray}
\frac{\partial \mathbb{P}^{a}}{\partial t}=-\gamma \mathbb{P}^{a}-\frac{v_{0}}{2}\nabla^{a}\rho-v_{0}\nabla_{b}\mathbb{Q}^{ab}.
\label{current}
\end{eqnarray}
Next, we need the evolution for the nematic order tensor field $\mathbb{Q}_{ab}(x,t)$.  This equation can be written in terms of the polarization field, and a  traceless three 
rank tensor field \cite{Fily2016}. However, equation for $\mathbb{Q}_{ab}(x,t)$  will not be used in the next calculations due to the approximation we are going to make 
below. Combining  both equations (\ref{cont}) and (\ref{current}) one is able to obtain 
\begin{eqnarray}
\frac{\partial^2\rho}{\partial t^2}+\gamma \frac{\partial \rho}{\partial t}=\frac{v^2_{0}}{2}\Delta_{g}\rho+v^2_{0}\nabla_{a}\nabla_{b}\mathbb{Q}^{ab},
\label{rho-Q}
\end{eqnarray}
where $\Delta_{g}$ is the so-called Laplace-Beltrami operator, given explicitly in generalized coordinates by
\begin{equation}\label{LaplaceBeltramiOP}
\Delta_{g}=\frac{1}{\sqrt{g(x)}}\frac{\partial}{\partial x^{a}}\sqrt{g(x)}g^{ab}(x)\frac{\partial}{\partial x^{b}}.
\end{equation}

With the probability density function, $\rho(x, t|x^{\prime})$, at hand, we look at the expectation values of physical observables, $\mathcal{O}\left(x\right)$, like the mean squared geodesic-displacement. The expectation values are defined in the standard fashion by
\begin{eqnarray}
\left<\mathcal{O}\left(x\left(t\right)\right)\right>=\int d^{2}x\sqrt{g}~\mathcal{O}\left(x\right)\rho(x, t|x^{\prime}).
\end{eqnarray}
It is noteworthy to mention that in the open euclidean space, where $g_{ab}=\delta_{ab}$, one is able to obtain an exact result for the MSGD, $\left<s^2\right>_{0}$, where the euclidean displacement is $s=\left|{\bf x}-{\bf x}_{0}\right|$. Indeed, by use of (\ref{rho-Q}) one is able to show that the action of the operator $\frac{d^2}{dt^2}+\gamma\frac{d}{dt}$ on $\left<s^{2}\right>_{0}$ is given by  $\frac{v_{0}^2}{2}\int d^{2}x s^2\left[\nabla^2\rho+2\partial_{a}\partial_{b}\mathbb{Q}^{ab}\right]$.
The integral can be evaluated by noticing that $\nabla^2 s^2=4$, $\partial_{a}\partial_{b}s^2=2\delta_{ab}$, and by use of the fact that $\mathbb{Q}^{ab}\delta_{ab}=0$. With 
these considerations it is straightforward to show that
\begin{eqnarray}
\frac{d^2}{dt^2}\left<s^{2}(t)\right>_{0}+\gamma\frac{d}{dt}\left<s^{2}(t)\right>_{0}=2v_{0}^2.
\label{eqeuclid}
\end{eqnarray}
The solution of this equation with initial conditions $\left<s^{2}(t)\right>_{0}=0$ and $\frac{d}{dt}\left<s^{2}(t)\right>_{0}=0$ at $t=0$ gives the exact expression \cite{SevillaPRE2014} 
\begin{eqnarray}\left<s^2(t)\right>_{0}=4D\left[t-\left(1-e^{-\gamma t}\right)/\gamma\right],
\end{eqnarray}
where $D=\frac{v_{0}^{2}}{2\gamma}$ is an effective diffusion constant. 

In the next section the polar approximation is considered. In this approximation the nematic order tensor field is assumed a fast variable and approximately homogeneous over the points of the curved surface. Thus second term in the rhs of Eq. (\ref{rho-Q}) will be neglected. The equation that turns out is called the telegrapher's equation.

\section{\label{SectIII} The polar approximation: The telegrapher equation}

To deal with the dynamics of the density function $\rho(x,t\vert x^{\prime})$, it is customarily to truncate the infinite hierarchy of equations of the last section, up to the dynamics of the polarization field $\mathbb{P}^{a}$ and disregard the contribution of higher multipole terms. Such a truncation may be justified on the following physical grounds: if the active particle is diffusing on the surface of a one-piece manifold, it is expected that as time passes, the density becomes uniform on the surface, i.e. $\rho(x,t\vert x^{\prime})\rightarrow \Omega_{\mathbb{M}}^{-1}$, where $\Omega_{\mathbb{M}}$ is the area of the manifold's surface. Thus, any inhomogeneity of the density at short times, is induced by the contribution of higher multipoles of the hierarchy. Each contribution is rapidly damped out with time, the higher the multipole in consideration the faster is damped out.
Notice that $\rho(x,t\vert x^{\prime})$, strictly speaking, depends on the initial position $x^{\prime}$ and that $\rho(x,t\vert x^{\prime})dx$ is the probability to find an 
active particle in the area $dx$ centered at $x$, at time $t$, when the particle started at $x^{\prime}$, at time $t=0$. 

Under these considerations we have that, in the polar approximation, $\rho(x,t\vert x^{\prime})$ satisfies the non-Euclidian version of the so-called Telegrapher's equation
\begin{equation}\label{TelegrapherEq}
\frac{\partial^2 \rho}{\partial t^2}
+\gamma\frac{\partial \rho}{\partial 
t}=\frac{v_{0}^{2}}{2}\Delta_{g}\rho.
\end{equation}
This equation has received much attention in different contexts \cite{DunkelPhysRep2009} that consider the flat geometry of space, however, to our knowledge, little or nothing 
has been said about the effects of intrinsic curvature on the transport properties described by such equation.

The formal solution to the Eq. \eqref{TelegrapherEq} can be found by expanding $\rho\left(x, t\vert x^{\prime}\right)$ in a complete set of eigenfunctions, 
$\left\{\Psi_{I}(x)\right\}$, of the Laplace-Beltrami operator, $-\Delta_{g}$. This method is valid for arbitrary one-piece (compact) manifolds $\mathbb{M}$, the Euclidean 
space $\mathbb{R}^{d}$ and manifolds that result form the direct product of these, namely, $\mathbb{M}\times\mathbb{R}$ (see table (\ref{hola})). Thus we have
\begin{eqnarray}
\rho\left(x,t\vert x^{\prime}\right)=\sum_{I}a_{I}\left(x^{\prime},t\right)\Psi_{I}\left(x\right), 
\label{pdfspan}
\end{eqnarray}
where the coefficients $a_{I}(x^{\prime},t)$ depends explicitly on the initial value $x^{\prime}$ and are given by (see appendix)
\begin{equation}\label{Coeff-a}
a_{I}(x^{\prime},t)=\sum_{i=\pm}\bar{K}^{(i)}_{I}\left(t\right)A_{i,I}\left(x^{\prime}\right).
\end{equation}
$\bar{K}^{\pm}_{I}\left(t\right)$ are the Green's functions,  
\begin{eqnarray}
\bar{K}^{\pm}_{I}\left(t\right)=\pm  \frac{\exp\left[\alpha_{\pm}\left(v^2_{0}\lambda_{I}^{2}/2\right)t\right]}{\alpha_{+}\left(v^2_{0}\lambda_{I}^{2}/2\right)-\alpha_{-}\left(v^2_{0}\lambda_{I}^{2}/2\right)},
\label{KernelGen}
\end{eqnarray}
that correspond to the two independent solutions of the characteristic equation associated to \eqref{TelegrapherEq}, which is equivalent to the second order differential equation of a damped harmonic oscillator. $A_{i,I}(x^{\prime})$ are functions of $x^{\prime}$ only and are determined from the initial data. The symbol $\lambda_{I}^{2}\ge0$, that have physical dimensions of Length$^{-2}$, denote the discrete set of eigenvalues of $-\Delta_{g}$, that correspond to the eigenfunctions $\Psi_{I}(x)$. The functions $\alpha_{\pm}\left(\omega\right)=-\frac{\gamma}{2}\pm\sqrt{\frac{\gamma^{2}}{4}-\omega^{2}}$ are derived in the appendix, where $\gamma$ is the inverse of the persistence time.

If the following initial data is chosen,
\begin{subequations}\label{IniConds}
\begin{align}
 \lim_{t\to 0}\rho(x,t\vert x^{\prime})&=\frac{1}{\sqrt{g}}\delta^{(2)}\left(x-x^{\prime}\right),\\
  \lim_{t\to 0}\frac{\partial \rho\left(x,t\right)}{\partial t}&=0,  \end{align}
\end{subequations}
namely, if a pulse on the surface starts to propagate with vanishing initial flux, the coefficients  $A_{\pm, I}\left(x^{\prime}\right)$ can be computed explicitly and after substitution in \eqref{Coeff-a} and some rearrangements in \eqref{pdfspan} we have
\begin{equation}
\rho(x,t\vert x^{\prime})=\sum_{I}G\left(\frac{\gamma t}{2},\frac{2v_{0}^{2}}{\gamma^{2}}\lambda_{I}^{2}\right)\, \Psi_{I}^{\dagger}(x^{\prime})\Psi_{I}(x),
\label{genres}
\end{equation}
where the function $G(v,w)$ is given explicitly as
\begin{equation}\label{G}
G(v,w)=e^{-v}\left[\cosh(v\sqrt{1-w})+\frac{\sinh(v\sqrt{1-w})}{\sqrt{1-w}}\right].
\end{equation}
This function embodies the time evolution that characterizes the telegrapher's equation, indeed, notice that for each eigenvalue $\lambda_{I}^{2},$ for which $\left(2 v_{0}^{2}/\gamma^{2}\right)\lambda_{I}^{2}>1$, $G$ shows an oscillatory behavior associated to the wave-like propagation originates by the second order time-derivative that appear in the telegrapher's equation.

In particular, the solution on the flat case is recovered in comparison with explicit solutions presented in refs. \cite{MorseFeshbachBook, PorraPRE1997}. Note, also, in this case that the particle density satisfies the boundary behavior  $\rho(x,t\vert x^{\prime})\rightarrow 0$ when $\vert x\vert\rightarrow\infty$.

\begin{table}[h]
\centering
\begin{tabular}{|l|l|l|l|}
\hline
 & &   &  \\ 
 {\bf Manifold}&{\bf Index} & {\bf Sum}& {\bf Eigenfunctions of $-\Delta_{g}$} \\
& & & \\
\hline
 &&&\\
$\mathbb{M}$ & $I$ & $\sum_{I}$  & $\Psi_{I}\left(x\right)$  \\
& & & \\
 \hline
&&&  \\
$\mathbb{R}^{d}$ & $p$ &$\int d^{d}p $    & $\exp\left({i p\cdot x}\right)/{\left(2\pi\right)^{d/2}}$   \\
& & & \\
 \hline
&&& \\
$\mathbb{M}\times \mathbb{R}$ &  $(I, p)$  & $\sum_{I}\int dp$  & $\Psi_{I}\left(x\right)\exp\left({ip z}\right)/\sqrt{2\pi}$  \\
& & & \\
\hline
\end{tabular}
\caption{{\small In this table we show the type of manifolds that we are considering and the corresponding identification of the index, sum and eigenfunctions.}}
\label{hola}
\end{table}

Notice that two characteristic length scales appear in this, so far, general analysis. One of these scales characterizes the persistence of active motion and we refer to it as 
the the persistence length, denoted with $L$ and given by the product of the swimming speed $v_{0}$ times the persistence time $\gamma^{-1}$, i.e. $L=v_{0}/\gamma$. The other 
length scale, characterizes the particular surface under consideration and can be chosen, without loss of generality, as the squared root of the inverse of the first positive 
eigenvalue, namely $R=1/\sqrt{\lambda_{1}^{2}}$ (this is warranted under the assumption of the compactness of the manifold, for which the zero eigenvalue is associated to the 
constant eigenfunction). Any of this two characteristic lengths can be picked out as the length scale in the system, and the ratio between them $R/L$, serves as a parameter 
that compares the effects of curvature to those of persistence, in the diffusion process of an active particle on the surface.

In the limit, $L\ll R$, we have that $G\left(\gamma t/2, 2\frac{v_{0}^{2}}{\gamma^{2}}\lambda_{I}^{2}\right)$ can be approximated by $ e^{-\lambda_{I}^{2}\, Dt}$, where 
$D=v^2_{0}/2\gamma$ is the well-known effective diffusion coefficient. In this limit the particle density is given by
\begin{equation}
\rho(x,t\vert x^{\prime})\simeq \sum_{I}\exp\left(-\lambda_{I}^{2}\, Dt \right) \Psi_{I}^{\dagger}(x^{\prime})\Psi_{I}(x),
\label{genres-diffusive}
\end{equation}
which corresponds to the formal solution of the diffusion equation in curved manifolds, i.e. $\partial\rho/\partial t=D\Delta_{g}\rho$. Notice that in the asymptotic limit 
$t\rightarrow\infty$, the dominating term corresponds to the constant function associated to the vanishing eigenvalue $I=0$, and therefore 
$\rho\rightarrow\vert\Psi_{0}\vert^{2}$. From the normalization condition we have that $\Psi_{0}^{\dagger}=\Psi_{0}=1\sqrt{\Omega_{\mathbb{M}}}.$  The mean squared distance 
from the initial position $x^{\prime}$, $\langle s^{2}(t)\rangle$ tends to the constant value $\int dx^{a} \left[d(x\vert x^{\prime})\right]^{2}/\Omega_{\mathbb{M}}$, where 
$d(x\vert x^{\prime})$ denotes the geodesic distance between $x$ and $x^{\prime}$.

In the opposite limit, $L\gg R$, the effects of persistence are important, and $G\left(\gamma t/2, 2v_{0}^{2}\lambda_{I}^{2}/\gamma^{2}\right)$ results oscillatory for each 
eigenvalue $\lambda_{I}>1/L$. In particular for $\gamma t\ll1$ one has that $G\left(\gamma t/2, 2v_{0}^{2}\lambda_{I}^{2}/\gamma^{2}\right)\simeq \cos\left(\lambda_{I}\, 
\frac{v_{0}}{\sqrt{2}}t\right)$ and therefore 
\begin{equation}
\rho(x,t\vert x^{\prime})\simeq \sum_{I}\cos\left(\lambda_{I}\, v_{0}t/\sqrt{2} \right) \Psi_{I}^{\dagger}(x^{\prime})\Psi_{I}(x),
\label{genres-wave}
\end{equation}
which now corresponds to a pulse that propagates on the surface of a compact, curved manifold, that started at $x^{\prime}$, which is a solution of the wave equation, $\partial^2\rho/\partial t^2=\frac{v^2_{0}}{2}\Delta_{g}\rho$. 

\section{\label{SectIV} Weak curvature approximation}

In this section, our goal is to determine an approximation for the probability density function, $\rho\left(x,t\right)$, in a neighborhood of a given point of the manifold. This approximation captures the first correction due to the effects of curvature which results linear in the Ricci curvature tensor $R_{ab}$ and would serve as a basis to the implementation of computing algorithms to find solutions on arbitrary surfaces.

The procedure used in this paper follows the same techniques originally used in the context of Quantum Field Theory in curved space, developed mainly by B. DeWitt 
\cite{DeWittBook} (see appendix), which goes in analogy with the standard perturbation theory in Quantum Mechanics. In addition, we use the such an approximation for $\rho(x, 
t)$, in order to compute an expression for the mean squared geodesic-displacement in the weak curvature regime.  

\subsection{The probability density function in the neighborhood of a given point on the surface}

For weakly curved surfaces, the pdf around the neighborhood of a given position $x^{\prime}$, can be approximated as the superposition of the continuous set of eigenfunctions $\Psi_{I}\left(x\right)=\exp\left(ip\cdot x\right)/\left(2\pi\right)$ with eigenvalue $p$ in the infinite interval $(-\infty,\infty)$. Since our interest is in providing the pdf around $x^{\prime}$, it is clear that for weakly enough curved surfaces such pdf can be approximated by its locally flat counterpart and therefore 
\begin{equation}
\rho\left(x,t\vert x^{\prime}\right)=\int \frac{d^{2}p}{\left(2\pi\right)}\, a(p;x,x^{\prime},t)\, e^{ip\cdot x},
\end{equation}
where the coefficient $a(p;x,x^{\prime}, t)$ is now given, in analogy with \eqref{Coeff-a}, by
\begin{equation}
a(p;x,x^{\prime},t)=\sum_{i=\pm}\bar{K}^{(i)}(p,x,t)A_{i}(p,x^{\prime}).
\end{equation}
In this approximation the effects of curvature are encoded in the Green functions $\bar{K}^{(i)}(p,x,t)$ only, i.e. curvature decouples explicitly from the complete set of eigenfunctions of the Laplace-Beltrami operator. The calculation of the Green functions follows the standard procedures used in the perturbation theory in Quantum Mechanics and are computed explicitly in the appendix, these are given by 
\begin{equation}
 \bar{K}^{\pm}\left(p,x,t\right)=\pm  \frac{g^{-1/4}\left(x\right)\exp\left\lbrace\alpha_{\pm}\left[H_{0}\left(p\right)\right]t\right\rbrace}{\alpha_{+}\left[H_{0}\left(p\right)\right]-\alpha_{-}\left[H_{0}\left(p\right)\right]},
 \label{Kernel}
 \end{equation}
where $\alpha_{\pm}(P)=-\frac{\gamma}{2}\pm\sqrt{\frac{\gamma^{2}}{4}-P}$ and
\begin{equation}
H_{0}\left(p\right)=\frac{v^2_{0}}{2}\left(p^2-\frac{R_{g}\left(x^{\prime}\right)}{6}\right),
\end{equation}
$R_{g}(x^{\prime})$ being the well-known scalar curvature (see appendix) evaluated at $x^{\prime}$. As before, $A_{i}(p,x^{\prime})$ are directly determined from the initial data \eqref{IniConds}, which lead to  
\begin{multline}
\rho(x,t\vert x^{\prime})\simeq \int \frac{d^{2}p}{(2\pi)}e^{ip\cdot (x-x^{\prime})}\, 
g^{-1/4}(x)g^{-1/4}(x^{\prime})\\
G\left[\frac{\gamma t}{2}, \frac{4}{\gamma^{2}}H_{0}\left(p\right)\right],
\end{multline}
with the function $G$ defined as in \eqref{G}. In order to obtain the linear curvature response we still need to expand the function $G(\frac{\gamma t}{2}, 
\frac{4}{\gamma^2}H_{0}\left(p\right))$ and $g^{1/4}(x)$ linearly in the curvature.

Finally, let $\tilde{\rho}(p,t)\equiv \left<e^{-ip\cdot y}\right>$ be the characteristic function , that is, the inverse Fourier transform of $\sqrt{g}~\rho\left(x,t\right)$, in the linear curvature response is given by
\begin{eqnarray}
\widetilde{\rho}(p,t)\simeq G\left(\gamma t/2, w\right)+\frac{4}{3}\left(\frac{v_{0}}{\gamma}\right)^{4}R_{ab}p^{a}p^{b}\frac{\partial^{2}G(\gamma t/2,w)}{\partial w^{2}},\nonumber\\
\end{eqnarray}
where we must evaluate at $w=\frac{2v_{0}^{2}}{\gamma^{2}}p^{2}$. Using the characteristic function one is able to compute all the moments of the distribution within the linear approximation.

\subsection{The mean squared geodesic-displacement: the weak curvature limit}

The linear curvature response in the mean-square displacement can be computed using the correlation function $\left<y_{a}(t)y_{b}(t)\right>=-\partial^2 \tilde{\rho}\left(p,t\right)/\partial p^{a}\partial p^{b}$, which is calculated  with a second derivative of the characteristic function. The structure of $\left<y_{a}(t)y_{b}(t)\right>$ is  inherited from the form of $\tilde{\rho}(p,t)$, that is, within the linear curvature approximation the correlation function displays a known flat expression plus  a first correction due to the curvature, $R_{ab}$,  which is multiplied by the function $f(v)$ given by
\begin{eqnarray}
f(v)=\frac{1}{4}\left[v^2-2v+\frac{3}{2}-\left(v+\frac{3}{2}\right)e^{-2v}\right]. 
\end{eqnarray}
One can notice that for dimension $d>2$, correlations between $y_{a}$ and $y_{b}$, for $a\neq b$, may occur depending  on the structure of the Ricci tensor, $R_{ab}$. However, 
in the dimension of our interest $d=2$, the Ricci tensor is proportional to the metric tensor, $g_{ab}$, and the scalar curvature, $R_{g}$. Furthermore, in $d=2$, the Riemann 
normal coordinates implies that $g_{ab}(x^{\prime})=\delta_{ab}$, where $x^{\prime}$ is the fiducial point. These considerations imply that 
$2\left<y_{a}(t)y_{b}(t)\right>=\left<s^2(t)\right>\delta_{ab}$, where the mean squared geodesic-displacement turns out to be 
\begin{eqnarray}
\left<s^2(t)\right>=\left<s^2(t)\right>_{0}-\frac{32}{3}\left(\frac{D}{\gamma}\right)^{2}f\left(\frac{\gamma t}{2}\right)R_{g}+\cdots,
\label{mainresult}
\end{eqnarray}
where the mean squared displacement, $\left<s^2(t)\right>_{0}=4D\left[t-\left(1-e^{-\gamma t}\right)/\gamma\right]$, on the flat space was computed previously by one of the 
authors \cite{SevillaPRE2014}.  
In what follows, we are going to determine the behaviors displayed by the mean squared geodesic-displacement  (\ref{mainresult}) in the limiting cases performing in the last 
section. 

For long time $t\gg 1/\gamma$ we have $f(\frac{\gamma t}{2})\simeq \gamma^2t^2/16$ which gives the mean squared geodesic-displacement, 
\begin{eqnarray}
\left<s^2(t)\right>\simeq 4Dt-\frac{2}{3}R_{g}(Dt)^2+\cdots,
\end{eqnarray}
for a Brownian particle in a curved manifold of scalar curvature $R_{g}$, where $D$ is the same effective diffusion coefficient defined above.  The first term corresponds to the standard diffusion regime found in an Euclidean space, whereas the second term corresponds to the first curvature correction that captures the way how curvature is affecting diffusion.  Further curvature corrections can be also determine using the same procedure \cite{CastroJStatMech2010}. 

For the short-time, $t\ll 1/\gamma$, we have $f(\frac{\gamma t}{2})\simeq \gamma^{4}t^{4}/192$, then one has the following curvature correction to the ballistic regime
\begin{eqnarray}
\left<s^2\right>\simeq v^2_{0}t^2-\left[\frac{1}{72}v^4_{0}t^4-\frac{1}{180}(v^4_{0}t^4)(\gamma t)+\cdots\right]R_{g}.\nonumber\\
\end{eqnarray}
The first terms corresponds to the standard ballistic regime found already in the flat spaces \cite{SevillaPRE2015}. The term $(v_{0}t)^4 R_{g}$ corresponds to the linear curvature response for the purely wave-like stochastic process i.e. when $\gamma=0$ in the equation (\ref{TelegrapherEq}). The next term $(v_{0}t)^{4}(\gamma t)R_{g}$ corresponds to the correction due to the way how the diffusion like stochastic process begin to compete with the wave-like stochastic process in the linear curvature regime.  

In a general local domain $\mathcal{D}\subset \mathbb{M}$, with $x^{\prime}\in \mathcal{D}$ and scalar curvature $R_{g}(x^{\prime})$, of the $2D$ manifold,  we claim that  Eq. (\ref{mainresult}) describes the crossover from ballistic motion on the curved manifold in the short-time regime to diffusive motion on the curved manifold in the asymptotic long-time limit. This is a generalization of the same crossover in Euclidean spaces discovered previously by one of the authors in \cite{SevillaPRE2014}. 

\section{\label{SectV} Active motion on the surface of the three-dimensional sphere $S^2$}
As an example of the direct application of the general theory given in the previous sections, we analyse in this one, the motion of an active particle moving on the surface of 
a sphere of radius $R$. This particular example has been discussed recently, by the use of numerical simulations, in Ref. \cite{ApazaPRE2017}. The kinematic state of the 2D 
active particle on the sphere can be described using equations (\ref{eq1}) specialized to the case of the sphere. In this case, there is a further simplification since one can 
choose a coordinate system where the metric tensor of $S^2$ is diagonal. 

The Riemannian geometry of the sphere $S^{2}$ is encoded in the metric tensor, which for the standard spherical coordinates $x^{\theta}=\theta$ and $x^{\varphi}=\varphi$, the  polar, and azimuthal angles respectively, has as elements the following ones: $g_{\theta\varphi}=g_{\varphi\theta}=0$, $g_{\theta\theta}=R^2$, and $g_{\varphi\varphi}=R^2\sin^2\theta$. In these coordinates the components of the swimming direction $\hat{\boldsymbol{v}}_{\text{swim}}$ are denoted with $v^{\theta}$ and $v^{\varphi}$, and the Christoffel symbols are explicitly given by $\Gamma\indices{^{a}_{\theta\theta}}=\Gamma\indices{^{\theta}_{\theta\varphi}}=\Gamma\indices{^{\varphi}_{\varphi\varphi}}=0$ with $a=\theta,\varphi$,
$\Gamma\indices{^{\varphi}_{\theta\varphi}}=\cot \theta$ and $\Gamma\indices{^{\theta}_{\varphi\varphi}}=-\sin\theta\cos\theta$
and the determinant of the metric tensor is $g=R^4\sin^2\theta$. With these particular values, we have that Eqs. \eqref{eq1b} can be written explicitly as
\begin{subequations}\label{vsphere}
\begin{align}
\frac{d}{dt}v^{\theta}_{\text{swim}}&=v_{0}\left(v^{\varphi}\right)^{2}\sin\theta\cos\theta-\sqrt{2\gamma}R\sin\theta\, v^{\varphi}\zeta[x(t)],\\
\frac{d}{dt}v^{\varphi}_{\text{swim}}&=-v_{0}v^{\theta}_{\text{swim}}v^{\varphi}_{\text{swim}}2\cot\theta+\sqrt{2\gamma}R\, v_{\theta}\, \zeta[x(t)],
\end{align}
\end{subequations}
where the state-dependent noise $\zeta[x(t)]$ is explicitly given by $\zeta_{1}(t) \cos\theta(t) \cos\varphi(t)+\zeta_{2}(t) \sin\theta(t) \sin\varphi(t)+\zeta_{3}(t) \cos\theta(t)$.
As a consequence of the constancy of the swimming speed, the $v^{a}_{\text{swim}}$'s are not independent and are related by $\left[R 
v^{\theta}_{\text{swim}}(t)\right]^{2}+\left[R\sin\theta\, v^{\varphi}_{\text{swim}}(t)\right]^{2}=1,$ therefore only one degree of fredom is needed. It is convenient to chose 
that degree of freedom as the swimming angle $\Theta(t)$, such that in terms of this, the components of the swimming direction are 
$v^{\theta}_{\text{swim}}(t)=\cos\Theta(t)/R$ and $
v^{\varphi}_{\text{swim}}(t)=\sin\Theta(t)/R\sin\theta(t)$. With these considerations we have that the Langevin equations that give the trajectories of an active particle on the surface of a sphere are given by 
\begin{subequations}
\label{eqsphere}
\begin{align}
\label{eqsphere1}
\frac{d}{dt}\theta(t)&=\frac{v_{0}}{R}\cos\Theta(t),\\
\frac{d}{dt}\varphi(t)&=\frac{v_{0}}{R}\frac{\sin\Theta(t)}{\sin\theta(t)},\label{eqsphere2}\\
\frac{d}{dt}\Theta(t)&=-\frac{v_{0}}{R}\sin\Theta(t)\, \cot\theta(t)+\sqrt{2\gamma}~\zeta[x(t)].
\label{eqsphere3}
\end{align}
\end{subequations}
In contrast with the corresponding equations for an active particle that diffuse in a two-dimesional Euclidean plane (see for intances those given in \cite{SevillaPRE2014}), two effects due to the sphere curvature can be identified, namely, the first term in the right hand side of Eq. \eqref{eqsphere3} that accounts for the intrinsic curvature of the sphere and secondly, the state dependent nature of the active fluctuations which leads to multiplicative noise.

The Fokker-Planck equation for the one-particle distribution function
\begin{eqnarray}
P(\theta, \varphi, \Theta, t)=\left<\frac{1}{\sqrt{g}}\delta\left(x^{a}-x^{a}(t)\right)\delta\left(\Theta-\Theta(t)\right)\right>,\nonumber
\end{eqnarray}
associated to Eqs. (\ref{eqsphere}) is given, after a straightforward calculation following the method in section \ref{SectII}, by
\begin{eqnarray}
\frac{\partial P}{\partial t}&=&\gamma\frac{\partial^2P}{\partial \Theta^2}-\frac{v_{0}\cos\Theta}{R\sin\theta}\frac{\partial}{\partial\theta}\left(\sin\theta P\right)\nonumber\\
&-&\frac{v_{0}\sin\Theta}{R\sin\theta}\frac{\partial}{\partial\varphi} P+
\frac{v_{0}}{R}\frac{\partial}{\partial\Theta}\left(\sin\Theta\cot\theta P\right),\nonumber\\
\end{eqnarray}
where the arguments of $P$ have been omited for the sake of writing. We now follow the same procedure used in Ref. \cite{SevillaPRE2014}, and we employ the following 
expansion,   
\begin{eqnarray}\label{ExpansionSphere}
P(\theta,\varphi, \Theta, t)=\sum_{n\in\mathbb{Z}}e^{in\Theta}e^{-\gamma n^{2}t}p_{n}\left(\theta,\varphi,t\right),
\end{eqnarray}
where the expansion coefficients $p_{n}(\theta,\varphi,t)$ satisfy the following hierarchy equations
\begin{eqnarray}
\frac{\partial {p}_{n}}{\partial t}=-\frac{v_{0}}{2R	}e^{-\gamma t}\sum_{\sigma=-1}^{1}e^{-2 \sigma\, \gamma n\, t}\hat{\ell}_{\sigma,n}\, p_{n+\sigma},\nonumber\\
\label{intermediateresult}
\end{eqnarray}
where the operators $\hat{\ell}_{\sigma,n}$, $\sigma=\pm1$, are given explicitly by 
\begin{equation}
\hat{\ell}_{\sigma,n} =\frac{1}{\sin\theta}\left(\frac{\partial }{\partial\theta}\sin\theta +\sigma i\frac{\partial}{\partial\varphi}\right)+\sigma n\cot\theta.
\end{equation}
The diffusion of free active particles on the sphere, is given by the exact solution of the hierarchical Eqs. (\ref{intermediateresult}), which is unknown in the most general case. We explore such a solution in the polar approximation as is discussed in the next section, leaving the analysis of the effects of higher Fourier modes in the expansion \eqref{ExpansionSphere}, to be presented elsewhere in a future communication. 

\subsection{Active motion on the sphere $S^2$: The polar approximation}

In the polar approximation, the first three Fourier modes, namely $p_{0}(\theta,\varphi,t)$ and $p_{\pm1}(\theta,\varphi,t)$, are retained in such a way  that the hierachical equations \eqref{intermediateresult} are reduced to a closed system of equations, after elimination of the modes $p_{\pm1}(\theta,\varphi,t)$ we have that $p_{0}(\theta,\varphi,t)$  satisfies the spherical teleprapher equation 
\begin{multline}
\frac{\partial^{2}}{\partial t^{2}}p_{0}+\gamma\frac{\partial}{\partial t}p_{0}=\frac{v_{0}^{2}}{2R^{2}}\left[\frac{1}{\sin\theta}\frac{\partial}{\partial\theta}\sin\theta\frac{\partial}{\partial\theta}+\right.\\
\left.\frac{1}{\sin^{2}\theta}\frac{\partial^{2}}{\partial\varphi^{2}}\right]p_{0}(\theta,\varphi,t),
\label{telsphere}
\end{multline}
where we have used that $\hat{\ell}_{-1,0}\hat{\ell}_{+1,-1}+\hat{\ell}_{+1,0}\hat{\ell}_{-1,+1}$ results into two times the Laplace-Beltrami operator, $\Delta_{g}$, in spherical coordinates, more precisely
\begin{multline}
\hat{\ell}_{-1,0}\hat{\ell}_{+1,-1}+\hat{\ell}_{+1,0}\hat{\ell}_{-1,+1}=2\times\\
\left[\frac{1}{\sin\theta}\frac{\partial}{\partial\theta}\sin\theta\frac{\partial}{\partial\theta}+\right.
\left.\frac{1}{\sin^{2}\theta}\frac{\partial^{2}}{\partial\varphi^{2}}\right].
\end{multline}
As before, we make  the identification $p_{0}(x,t)\equiv \rho(x,t\vert x^{\prime})$. 

As is well-known, the eigenfunctions of the Laplace-Beltrami operator on the sphere correspond to those given by the spherical harmonics $Y_{l}^{m}(\theta,\varphi)$ with eigenvalues $l(l+1)/R^2$, with $l=0,1, \ldots$ and $m=-l, \cdots, l$. If the initial probability distribution corresponds to a pulse with zero velocity in the north pole, the azimuthal invariance allows us to write the solution in the following manner
\begin{equation}
\rho(\theta,\varphi,t)= \sum_{l=0}^{\infty}\frac{2l+1}{4\pi}\, G\left[\frac{\gamma t}{2},2\frac{v_{0}^{2}}{\gamma^{2}}\frac{l(l+1)}{R^{2}}\right]\, P_{l}(\cos\theta),
\label{pdfsphere}
\end{equation}
where $P_{l}(\cos\theta)$ denotes the Legendre polynomial of degree $l$. Notice the explicit appearance of the ratio $R/L=R/(v_{0}/\gamma)$, which measures the competence between the effects of curvature and the effects of persistence. If the persistence length is much more smaller than the curvature radius, $L/R\rightarrow 0$, the well-known solution of diffusion on the sphere,
\begin{equation}
\rho(\theta,\varphi,t)= \sum_{l=0}^{\infty}\frac{2l+1}{4\pi}\, e^{-D\, l(l+1)t/R^{2}}\, P_{l}(\cos\theta),
\end{equation}
is recovered, with the effective diffusion constant $D=v^{2}/2\gamma.$

\subsubsection{Mean squared geodesic-displacement on $S^2$}
Coincident with the initial data previously chosen, the geodesic displacement $s(t)$ is $R\theta(t)$ and correspondingly the mean square geodesic-displacement, $\left\langle s^2(t)\right\rangle$ is given by $R^{2}\langle\theta^{2}(t)\rangle$. A differential equation for the mean square geodesic-displacement can be obtained directly from the Eq. \eqref{telsphere}, namely 
\begin{multline}
\frac{d^{2}\left\langle s^{2}(t)\right\rangle}{dt^{2}}+\gamma\frac{d \langle s^{2}(t)\rangle}{dt}=v_{0}^{2}\times\\
\left\lbrace 1+\left\langle\frac{s(t)}{R}\cot\left[\frac{s(t)}{R}\right]\right\rangle\right\rbrace.
\label{diffs2}
\end{multline}
In contrast with its counterpart in the two dimensional Euclidean manifold (\ref{eqeuclid}), the last equation is not closed in $\left\langle s^{2}(t)\right\rangle$ but coupled in a high nonlinear way with higher moments of $s(t)$, however, by the use of the Taylor expansion of $z\cot z=\sum_{n}(-1)^{n}2^{2n}B_{2n}z^{2n}/(2n)!$, where $B_{n}$ are the Bernoulli numbers, we have in the limit of weak curvature, i.e. $v_{0}/\gamma\ll R$, that  right hand side of Eq. (\ref{diffs2}) is
$2v_{0}^{2}\left[1-\frac{1}{6R^{2}}\left\langle s^{2}\right\rangle-\ldots\right]$. By retaining only the first correction proportional to $R^{-2}$ and recalling that the Ricci 
scalar curvature is $R_{g}=2/R^2$, the solution to the last equation coincides with the mean square geodesic-displacement given in Eq. \eqref{mainresult}. 

The exact time-dependence of $\left\langle s^{2}(t)\right\rangle$ can be obtained by the use of \eqref{pdfsphere}, which is given explicitly by
\begin{multline}
\left\langle s^2(t)\right\rangle=\frac{R^2}{2}\sum_{l=0}^{\infty}(2l+1)g_{\theta^2}\left(l\right)G\left[\frac{\gamma t}{2}, \frac{2 L^{2}}{R^2}l(l+1)\right],
\label{MSDsphere}
\end{multline}  
where $g_{\theta^2}\left(l\right)$ denotes the projection of $\theta^2$ onto the Legendre polynomial of degree $l$, i.e.
\begin{equation}
g_{\theta^{2}}(l)=\int_{0}^{\pi}d\theta\, \sin\theta\, P_{l}(\cos\theta)\, \theta^{2},
\end{equation}
and whose explicit dependence on $l$ has been given in Ref. \cite{CastroJChemPhys2014}. As before, $L$ denotes the persistent length.
\begin{figure}[h!]
\includegraphics[width=\columnwidth]{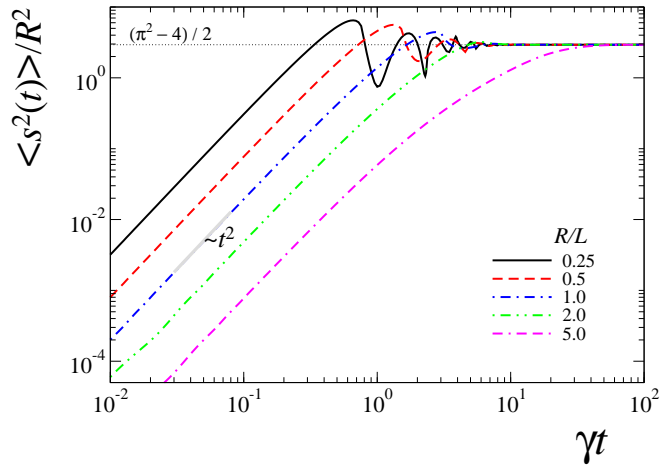}
\caption{(Color online) Time dependence of the dimensionless mean squared geodesic-displacement $\langle s^{2}(r)\rangle/R^{2}=\langle\theta^{2}(t)\rangle$ for different values of the ratio of the sphere radius $R$ to the persistent length $L$, namely, $R/L=0.25$, 0.5, 1.0, 2.0 and 5.0. The dotted-line marks the asymptotic value $\left\langle\theta^{2}(t)\right\rangle=(\pi^{2}-4)/2$ that characterizes the uniform distribution on the whole sphere, and the thick-gray line marks the $t^{2}$ dependence.} 
\label{FigMSD}
\end{figure}

In the Figure (\ref{FigMSD}) the time dependence of the mean squared geodesic-displacement given in Eq. \eqref{MSDsphere} is shown for some particular values of the ratio 
$R/L$. As can be observed in the same figure, an active particle confined to the sphere exhibits two conspicuously different behaviors whenever $R/L$ is larger or smaller than 
2. On the one hand, for $R/L\le2$, the mean squared geodesic-displacement starts growing quadratically with time (ballistic regime) in contrast to the linear grow for standard 
diffusion (attained in the $R/L\gg1$ regime). The explicit time dependence of $\left\langle s^{2}(t)\right\rangle$ in the ballistic regime can be written as 
$v_{\text{eff}}^{2}t^{2}$, with $v_{\textrm{eff}}$ an effective swimming speed defined through $v_{\text{eff}}^{2}=\eta^{2}\, v_{0}^{2}/8$, where the constant 
$\eta^{2}=-\sum_{l=0}^{\infty}l(l+1)(2l+1)g_{\theta^{2}}(l)>0$. 

In addition, the mean squared geodesic-displacement reaches the asymptotic value $(\pi^{2}-4)/2$ non-monotonically exhibiting oscillations, contrary to the case $R/L>2$ for which such behavior is monotonic. These oscillations have been pointed out in the numerical analysis of Ref. \citet{ApazaPRE2017}, though in there, the authors consider translational fluctuation in addition to rotational ones. The physical meaning of such oscillatory behavior is clear: The initial pulse in the north pole of the sphere starts to propagate with speed $v_{0}$ in all directions forming a sharp ring that sweeps the sphere surface (and in general the surface of a compact manifold) a number of times that depends on the ratio $R/L$. Indeed, this is confirmed by the estimation of the time at which the first maximum of the oscillations appears, which roughly corresponds to the time at which the particles reach the south pole of the sphere, $t_{\text{south}}$, a simple calculation leads to $t_{\text{south}}\approx(2\sqrt{2}\pi/\sqrt{\eta^{2}})\, R/v_{0}$ that gives $\gamma t_{\text{south}}\approx0.5878$ for $R/L=0.25$. As time passes the ring becomes thicker due to fluctuations, persistence effects become negligible and the distribution turns uniform in the asymptotic limit. For large values of the $R/L$ the effects of persistence are damped out leading to a standard diffusive regime.

\section{\label{SectVI} Concluding remarks and perspectives}

In this paper, we have analyzed the diffusion of non-interacting active particles confined to move on a curved surface. On the one hand, Langevin equations that consider 
explicitly the effects of curvature and active motion are provided. By the use of the corresponding Fokker-Planck equation, we built the standard stochastic hydrodynamic 
hierarchy of equations that couple the particle density $\rho$, the polarization field $\mathbb{P}^{a}$, the nematic tensor $\mathbb{Q}_{ab}$, etc. Of particular importance, 
the commonly used polar approximation, was considered. Such approximation consists in truncating the hierarchy of equations retaining only up to the polarization field and 
disregarding higher order tensors. The approximation is valid in the long-time regime when the nematic tensor can be considered as a fast variable and homogeneous over the 
surface. As consequence of the approximation, it was shown that the conserved particle density obeys a generalization of the Telegrapher's equation in curved surfaces, where 
the Laplace operator in Euclidean space, is replaced by the corresponding Laplace-Beltrami operator that considers the intrinsic curvature of the surface.

The main consequences of the generalization of the Telegrapher's equation to curved manifolds were discussed. On the one hand, a general solution is given for compact manifolds in terms of a expansion on the discrete set of eigenfunctions of the Laplace-Beltrami operator. In the short-time regime, the provided solution corresponds to the solution of the wave-equation in curved surfaces that characterizes wave-like propagation. In the weak-curvature limit, such propagation is reminiscent of the propagation in the plane, i.e. with propagation speed $v_{0}$ and wake effects are markedly observed. Interestingly, however, for arbitrary values of curvature, the signal propagation is realized at an effective speed that depends on the surface curvature. In a local domain, we have studied the effects of curvature in the probability density function as well as in the mean squared geodesic displacement. 

If initial data corresponding to a pulse with vanishing current is chosen, the pulse turns with time into a ring-like structure that propagates with the effective speed and for compact surfaces and for small enough ratios $R/L$, the ring structure recurs with time that lead to oscillations in the mean squared geodesic-displacement. This is clearly exhibited in section \ref{SectV} in the case of the sphere.

In the regime for which the persistence length is much smaller than the curvature that characterizes the surface, $L/R\rightarrow 0$, the solution is close to the solution of 
the diffusion equation on curved manifolds with an effective diffusion constant $D=v^{2}_{0}/2\gamma$. In this regime the dispersive term dominates over the inertial one. It 
is shown, particularly, that the MSGD in the diffusive regime coincides with that previously obtained for passive Brownian particles in curved space 
\cite{CastroJStatMech2010}. In the persistent regime, it is shown how the MSGD has a ballistic behavior, in particular, we provide corrections to this behavior when the 
effects of curvature and diffusive effects begin to be relevant. 

The results presented in this study can be extended along several directions. Among these are: the realization of a systematic study of the dynamics of active particles in a 
sphere beyond the polar approximation; the inclusion of passive fluctuations on the translational degree of freedom of the model can be treated in the same way to obtain 
analytical results. In particular, this situation could be subjected to an experimental scrutiny as has been the case for passive Brownian particles on the sphere 
\cite{ZhongJPhysChemC2017}. Another natural extension can be developed to include the effects of curvature in continuous mean field models similar to those of Toner and Tu 
hydrodynamic equations. Also, we can derive the hydrodynamic equations of Brownian particles with alignment interaction in curved space. Finally, the methods and results 
proposed in the present work allow us to propose a step further to develop simulation algorithms to study active particles with alignment interaction in different surfaces 
such as ellipsoids, tori, catenoid, etc.

\begin{acknowledgments}
FJS acknowledges support from DGAPA-UNAM through the grant PAPIIT-IN114717. PCV acknowledges financial support by CONACyT Grant No. 237425 and PROFOCIE-UNACH 2016 and 2017. 
\end{acknowledgments}

\appendix

\section{\label{appendix1}Langevin equations for active particles moving on surfaces}

The starting point in the derivation of Eqs. \eqref{eq1} is the following pair of equations 
\begin{subequations}
\label{Underdamped}
\begin{align}
\ddot{x}^{c}+\Gamma\indices{^{c}_{ab}}\dot{x}^{a}\dot{x}^{b}&=\frac{\mu}{m}{\bf e}^{c}\cdot \left[-\boldsymbol{e}_{b}\dot{x}^{b}+v_{0}\hat{\boldsymbol{v}}_{\text{swim}}(t)\right],\label{2order}\\
\frac{d}{dt}\hat{\boldsymbol{v}}_{\text{swim}}(t)&=\sqrt{2\gamma}\, \boldsymbol{\zeta}(t)\times\hat{\boldsymbol{v}}_{\text{swim}}(t),\label{SwimDynamics}
\end{align}
\end{subequations}
where $a,b,c=1,2$, $m$ is the mass of the particle and $\mu$ the dragging coefficient of the friction force exerted by the surface on the particle. $\lbrace {\bf e}_{a} 
\rbrace$ form a set of linearly independent local vectors at the position of the particle on the surface. Equation \eqref{2order} corresponds to the equation of motion of 
particle of mass $m$ moving on a surface \cite{CastroJChemPhys2014} subject to a linear friction force (first term in squared parenthesis) and to self-propulsion force-like 
(second term). Equation \eqref{SwimDynamics}, on the other hand, accounts for the internal dynamics of the self-propelling ``swimming force", that accounts for the stochastic 
rotations of $\hat{\boldsymbol{v}}_{\text{swim}}(t)$ in the tangent plane on the surface at the position of the particle, and $\boldsymbol{\zeta}(t)$ is a three-dimensional 
vector whose entries correspond to Gaussian white noise with zero mean and unit variance.  

We consider the overdamped limit, that is, the limit for which inertial effects can be neglected, consequently the left hand side of equation (\ref{2order}) is identical to 
zero which directly leads to \eqref{eq1a}. We also ignore any possible Brownian contribution due to external-- thermal-- fluctuations. 

The change in time of $\hat{\boldsymbol{v}}_{\text{swim}}(t)$ involves the change in time of local coordinate system, and therefore the change of the local basis $\lbrace{\bf 
e}_{a}\rbrace$. The variation of the local coordinate system upon small movements on the surface is accounted by the Weingarten-Gauss formulas, $\partial_{b}{\bf 
e}_{a}-\Gamma\indices{^{c}_{ba}}{\bf e}_{c}=-K_{ab}{\bf N}$, where $K_{ab}$ denotes  the components of the second fundamental form of the surface, namely, $K_{ab}={\bf 
e}_{a}\cdot \partial_{b}{\bf N}$. 
Thus the lhs of (\ref{SwimDynamics}) can be written as
\begin{multline}
\frac{d}{dt}\hat{\boldsymbol{v}}_{\text{swim}}(t)=\frac{d}{dt}v^{a}_{\text{swim}}\, {\bf e}_{a}+\\
v^{a}_{\text{swim}}\left(\Gamma\indices{^{c}_{ba}}{\bf e}_{c}-K_{ab}{\bf N}\right)\frac{dx^{b}}{dt}.
\label{actived}
\end{multline}
Since the dynamics occurs in the tangent plane we project equation \eqref{SwimDynamics} into it, thus the left hand side becomes
\begin{equation}
\frac{dv_{\text{swim}}^{a}}{dt}+v_{0}v^{a}_{\text{swim}}v^{b}_{\text{swim}}\Gamma\indices{^{c}_{ba}},
\end{equation}
where \eqref{eq1a} has been used to replace $dx^{b}/dt.$ The projection of the right hand side of the equation, $\sqrt{2\gamma}\boldsymbol{\zeta}\times \hat{\boldsymbol{v}}_{\text{swim}}\cdot{\bf e}_{d}$ can be written as
\begin{equation}
\sqrt{2\gamma}\,\sqrt{g}\, v_{\text{swim}}^{f}\, \epsilon_{fd}\, \boldsymbol{\zeta}\cdot{\bf N}.
\end{equation}
These considerations lead straightforward to equations (\ref{eq1}). 

\section{Green functions for the curved Telegrapher's equation}

\begin{figure}[h!]
\begin{center}
\includegraphics[width=0.35\textwidth, clip=true]{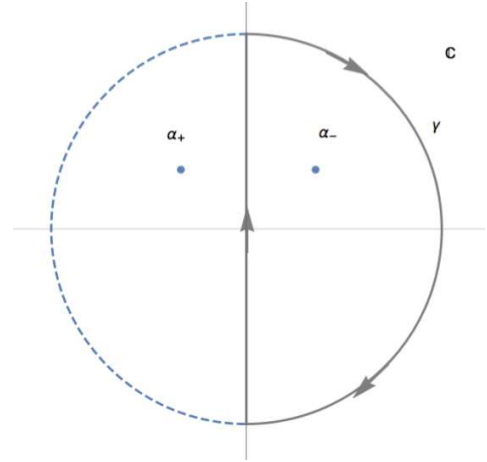}
\caption{It is shown one of the two possible complex integration contour. These contours are symmetric respect to the imaginary axis and they enclose the poles $\alpha_{+}$ and $\alpha_{-}$, respectively. }

\label{fig55}
\end{center}
\end{figure}

In this section, our goal is to determined the Green function $K\left(x,x^{\prime},t\right)$ that satisfies the equation 
\begin{eqnarray}\left(\frac{\partial^2}{\partial t^2}+\gamma\frac{\partial}{\partial t}-\frac{v^2_{0}}{2}\Delta_{g}\right)K\left(x,x^{\prime},t\right)=\frac{1}{\sqrt{g}}\delta\left(x-x^{\prime}\right)\delta\left(t\right).\nonumber\\
\label{GreenEqurr}
\end{eqnarray}
Now, in order to find a formal solution for $K\left(x,x^{\prime},t\right)$ we assume the existence of a complete set of eigenfunctions $\left\{\Psi_{I}\right\}$ and 
corresponding eigenvalues $-\lambda_{I}$ of the Laplace-Beltrami operator $\Delta_{g}$ \cite{ChavelBook}. Using the completness relation 
$\delta\left(x-x^{\prime}\right)/\sqrt{g}=\sum_{I}\Psi^{\dagger}_{I}\left(x^{\prime}\right)\Psi_{I}\left(x\right)$ and a Fourier decomposition in the time variable 
$K\left(x,x^{\prime},t\right)=\int \frac{dE}{2\pi} e^{i E t}K\left(x,x^{\prime},E\right)$ one can express the Green function as follows
\begin{eqnarray}
K\left(x,x^{\prime},t\right)=\sum_{I}\Psi^{\dagger}_{I}\left(x^{\prime}\right)\Psi_{I}\left(x\right) \oint_{\gamma}  \frac{dz}{2\pi i}\frac{e^{z t}}{z^2+\gamma z+\frac{v^2_{0}}{2}\lambda_{I}},\nonumber\\
\end{eqnarray}
where integration in the $E$ variable has been replacement by an equivalent complex integration on the semi-circle contour $\gamma$ (see figure (\ref{fig55})). We can identify the poles  
\begin{eqnarray}
\alpha_{\pm}\left(\omega\right)=-\frac{\gamma}{2}\pm\sqrt{\frac{\gamma^2}{4}-\omega},
\label{poles}
\end{eqnarray}
where $\omega=\frac{v^2_{0}}{2}\lambda_{I}$. 

The complex integration gives the two independent solutions
\begin{eqnarray}
\bar{K}^{\left(\pm\right)}\left(x,x^{\prime},t\right)=\pm \sum_{I}\frac{\Psi^{\dagger}_{I}\left(x^{\prime}\right)\Psi_{I}\left(x\right)e^{\alpha_{\pm}\left(v^2_{0}\lambda_{I}/2\right)t}}{\alpha_{+}\left(v^2_{0}\lambda_{I}/2\right)-\alpha_{-}\left(v^2_{0}\lambda_{I}/2\right)}.\nonumber\\
\label{greenf}
\end{eqnarray}
Using these Green's functions the  pdf can be obtained spanned in the Green's functions as follows 
\begin{eqnarray}
\rho\left(x,t\right)=\int d^{d}y\sqrt{g}\sum_{i=+,-}\left[K^{(i)}\left(x,y,t\right)A_{i}\left(y, x^{\prime}\right)\right].\nonumber\\
\label{pdfspan-c}
\end{eqnarray}
Now, we substitute $A_{i}\left(y, x^{\prime}\right)=\sum_{I}A_{i, I}\left(x^{\prime}\right)\Psi_{I}(y)$ and (\ref{greenf}) into (\ref{pdfspan-c}). In addition, we use the orthogonal relation $\int d^{d}y\sqrt{g}\Psi_{I}^{\dagger}\left(y\right)\Psi_{I}\left(y\right)=\delta_{II^{\prime}}$. All these considerations allow us to prove Eq. (\ref{pdfspan}). 

\subsection{Green functions for the curved Telegrapher's equation in the weak curvature regime}

By the use of the same methods originally implemented in the context of Quatum Field Theory on curved spaces \cite{DeWittBook,ParkerBook}, we now determine the Green's 
function, $K(x,x^{\prime},t)$, in the weak curvature regime. We first performed a change 
$K\left(x,x^{\prime},t\right)=g^{-1/4}\left(x\right)\bar{K}\left(x,x^{\prime},t\right)g^{-1/4}\left(x^{\prime}\right)$. In addition, the term $1/\sqrt{g}$ that multiplies by 
the Dirac delta appearing in the corresponding Green equation (\ref{GreenEqurr}) is separated as $g^{-1/4}\left(x\right)g^{-1/4}\left(x^{\prime}\right)$. Thus, using these 
changes of variables the resulting Green equation can be rewritten as
\begin{equation}\label{GreenEc}
\left[\frac{\partial^2}{\partial t^2}+\gamma\frac{\partial}{\partial t}+\hat{H}\right]\bar{K}\left(x, x^{\prime}, t\right)=\delta\left(x-x^{\prime}\right)\delta\left(t\right),
\end{equation}
where $\hat{H}(x)$ is defined in terms of the Laplace-Beltrami operator as $-\frac{v^2_{0}}{2}\, g^{1/4}(x)\, \Delta_{g}\, g^{-1/4}(x)$, which after use of the explicit 
definition of $\Delta_{g}$ given in Eq. \eqref{LaplaceBeltramiOP}, $\hat{H}(x)$ can be rewritten in terms of the operator $\hat{p}_{a}=-i\partial/\partial x^{a}$ as 
\cite{OConnorPhDThesis1985}
\begin{equation}
\hat{H}=\frac{v^2_{0}}{2}\left[\delta^{ab}\hat{p}_{a}\hat{p}_{b}+\hat{p}_{a}\left(\left(g^{ab}-\delta^{ab}\right)\hat{p}_{b}\cdot\right)+V\left(x\right)\right]
\end{equation}
with 
\begin{equation}
V\left(x\right)=-g^{-1/4}(x)\frac{\partial}{\partial x^{a}}\left[\sqrt{g(x)}\, g^{ab}(x)\frac{ \partial}{\partial x^{b}}g^{-1/4}(x)\right].
\end{equation}

Before attempting to do any approximation let us take the Fourier transform in the time variable of Eq. \eqref{GreenEc}. Let $\bar{K}\left(x,x^{\prime}, E\right)$ be the Fourier transform of the Green function where $E$ is the conjugate Fourier variable of time, then, Eq. \eqref{GreenEc} can be written as
\begin{eqnarray}
\left(-E^2+i\gamma E+\hat{H}\right)\bar{K}\left(x,x^{\prime},E\right)=\delta(x-x^{\prime}).
\end{eqnarray}
Thus the Green function can be written as the matrix elements of the resolvent operator $\hat{K}=\left[-E^2+i\gamma E+\hat{H}\right]^{-1}$ as 
\begin{equation}
\bar{K}\left(x, x^{\prime},t\right)=\left<x\left|\hat{K}\right|x^{\prime}\right>.
\end{equation}

We now take the advantage of the spatial covariance of the Laplace-Beltrami operator in the Telegrapher's equation, in order to use Riemann normal coordinates (RNC) around the 
point $x^{\prime}$. This coordinate frame is particularly useful in the case of weak curvature regime since one the following expansions are valid \cite{EisenhartBook} for the 
metric tensor $g_{ab}(x)$ and $\sqrt{g(x)}$
\begin{align}
g_{ab}(x)&=\delta_{ab}+\frac{1}{3}R_{acdb}(x^{\prime})\, (x-x^{\prime})^{c}(x-x^{\prime})^{d}+\ldots,\\
\sqrt{g(x)}&=1-\frac{1}{6}R_{ab}(x^{\prime})\, (x-x^{\prime})^{a}(x-x^{\prime})^{b}+\ldots,
\end{align}
where standardly the Riemann curvature tensor $R_{abcd}=g_{af}R^{f}_{bcd}$, $R^{a}_{bcd}=\partial_{c}\Gamma_{bd}^{a}-\partial_{d}\Gamma_{bd}^{a}+\Gamma^{a}_{cs}\Gamma^{s}_{bd}-\Gamma^{a}_{ds}\Gamma^{s}_{bc}$ and $R_{ab}=R^{c}_{acb}=g^{cd}R_{acbd}$ is the Rici tensor.

Using these expressions and the corresponding ones for the inverse metric tensor $g^{ab}(x)$ as well as for the determinant of the metric $g(x)$, one has the following approximation for the $\hat{H}$ 
\begin{eqnarray}
\hat{H}=\hat{H}_{0}+\hat{H}_{I},
\end{eqnarray}
where the unperturbed part $\hat{H}_{0}$ is given by 
\begin{equation}
\hat{H}_{0}=\frac{v^{2}_{0}}{2}\left[\delta^{ab}\hat{p}_{a}\hat{p}_{b}-\frac{R_{g}}{6}\right],
\end{equation}
being $R_{g}=g^{ab}R_{ab}$ the scalar curvature and the perturbing part 
\begin{equation}
\hat{H}_{I}=-\frac{1}{3}R\indices{_{acdb}}\hat{p}^{a}y^{c}y^{d}\hat{p}^{b}. 
\end{equation}

The standard perturbation theory used in Quantum Mechanics allows to write the resolvent operator as the following expansion
\begin{eqnarray}
\hat{K}=\hat{K}_{0}-\hat{K}_{0}\hat{H}_{I}\hat{K}_{0}+\cdots,
\end{eqnarray}
where the unperturbed resolvent operator $\hat{K}_{0}$ is defined by 
\begin{equation}
\hat{K}_{0}=\left[-E^2+i\gamma E+\hat{H}_{0}\right]^{-1},
\end{equation}
and can be referred to as the ``free" resolvent operator. Due to the antisymmetric nature of the Riemann tensor (see for instance \cite{CastroPRB2017}) it can be proved that 
$\left<x\left|\hat{K}_{0}\hat{H}_{I}\hat{K}_{0}\right|x^{\prime}\right>=0$ and therefore, to the lowest nontrivial approximation in the curvature, only the free resolvent 
operator contributes. With this considerations the Green function can be written as
\begin{multline}\label{GreenFuncApB}
\bar{K}\left(x,x^{\prime},t\right)=\int \frac{d^{d}p}{\left(2\pi\right)^{d}}e^{i p\cdot (x-x^{\prime})}\times\\
\oint_{\gamma} \frac{dz}{2\pi i}\frac{e^{z t}}{z^2+\gamma z+H_{0}\left(p\right)},
\end{multline}
where the complete set of eigenstates of $\hat{p}_{a}$, $\{\left|p_{a}\right>\}$, for which $\langle x^{a}\vert p_{b}\rangle=\delta_{ab}e^{ix^{a}p_{a}}$ and $H_{0}\left(p\right)=\frac{v^{2}_{0}}{2}\left(p^2-\frac{R_{g}}{6}\right)$, has been used. The dependence in time is recovered by the integration on the complex plane $z$, along the semi-circle contour $\gamma$ (see figure (\ref{fig55})), form which we obtain
\begin{equation}
\bar{K}^{\pm}\left(x,x^{\prime},t\right)=\pm\int \frac{d^{d}p}{\left(2\pi\right)^{d}}\frac{e^{ip\cdot (x-x^{\prime})+\alpha_{\pm}\left(H_{0}\left(p\right)\right)t}}{\alpha_{+}\left(H_{0}\left(p\right)\right)-\alpha_{-}\left(H_{0}\left(p\right)\right)}.
\end{equation}
where $\alpha_{\pm}\left(P\right)=-\frac{\gamma}{2}\pm\sqrt{\frac{\gamma^{2}}{2}-P}$ correspond to the poles of the integrand in Eq. \eqref{GreenFuncApB}

\end{document}